\documentclass{article}
\usepackage{graphicx} 
\usepackage{authblk}

\usepackage{subcaption}
\usepackage{xcolor}
\usepackage[numbers,super]{natbib}
\usepackage{hyperref}
\usepackage{amsmath}
\usepackage{comment}
\usepackage{tcolorbox,amsfonts,amssymb,caption}
\usepackage{float}
\floatstyle{plain}
\newfloat{boxfloat}{htbp}{lob}
\floatname{boxfloat}{Box}
\title{Generalized Maximum Entropy: When and Why you need it}
\author[1,2$\ast$]{Giuseppe M. Ferro}
\author[3,4,$\ast$]{Edwin T. Pos}
\author[5,$\ast$]{Andrea Somazzi}

\affil[1]{Center for Statistics \& Machine Learning, Princeton University, Princeton, USA}
\affil[2]{Department of Ecology and Evolutionary Biology, Princeton University, Princeton, NJ 08544}
\affil[3]{Quantitative Biodiversity Dynamics, Ecology and Biodiversity, Utrecht University Botanic Gardens, Utrecht University, Padualaan 8, Utrecht, 3584 CH, The Netherlands}
\affil[4]{Botanic Gardens, Utrecht University, Budapestlaan 17, Utrecht, 3584 CD, The Netherlands}
\affil[5]{IMT School for Advanced Studies, Piazza S. Francesco 19, 55100 Lucca, Italy}

\affil[$\ast$]{These authors contributed equally.}

\date{}

\begin{document}

\maketitle
{
\vspace{-1.2cm}
\begin{center}\footnotesize \textbf{Authors’ Contributions}\end{center}
\vspace{-0.3cm}
\footnotesize
G.M.F., A.S. and E.T.P. jointly conceived the study and designed the research.
A.S. carried out the simulations and analysed the results.
G.M.F. and E.T.P. developed the ecological and economic case studies and connected
the framework to real-world examples. All authors contributed to writing the
manuscript and approved the final version.
}

\normalsize
\begin{abstract}
The classical Maximum--Entropy Principle (MEP) based on Shannon entropy is widely used to construct least–biased probability distributions from partial information.  However, the Shore--Johnson axioms that single out the Shannon functional hinge on \emph{strong system independence}---an assumption often
violated in real‐world, strongly correlated systems.  
We provide a self-contained guide to \emph{when and why} practitioners should abandon the Shannon form in favour of the one-parameter Uffink--Jizba--Korbel (UJK)
family of generalized entropies.  After reviewing the Shore and Johnson axioms from an applied perspective, we recall the most commonly used entropy functionals and locate them within the UJK family. The need for generalized entropies is made clear with two applications, one routed in economics and the other in ecology. A simple mathematical model worked out in detail shows the power of generalized maximum entropy approaches in dealing with cases where strong system independence does not hold.
We conclude with practical guidelines for choosing an entropy measure and reporting results so that analyses remain transparent and reproducible.
\end{abstract}

\section{Introduction}
In several disciplines of science, ranging from economics and ecology to physics and the social sciences, entropy measures are considered crucial for describing and modeling complex systems \cite{golan2022information}. One particular application concerns deriving the most likely probability distribution related to a pattern of interest based on prior information \cite{jaynes1957information1,jaynes1957information2} (see Box \ref{box:maxent}). These patterns can be as varied as the position and momentum of gas molecules, the concentration of wealth among countries or age categories or even the distribution of individuals over species. In each case, the key step is to connect macro-state variables—coarse-grained descriptors of the system—to the underlying distribution of micro-states. This connection allows us either to infer causal mechanisms or to use the inferred distribution as input for modeling the system as a whole\cite{cover1999elements}. The framework of entropy maximization allows for a more objective understanding based on a top-down approach using only available data instead of incorporating prior beliefs based on assumptions (as is the case for many parametric methods). The Maximum (information) Entropy Principle (hereafter referred to as MEP), stemming from the fields of statistical mechanics and information theory, thus allows us to derive the most likely probability distribution that best represent our current state of knowledge. It does so by maximizing the information entropy of a distribution of interest while respecting certain known constraints, expressed often in the form of expected values using the principle of Lagrange multipliers, that reflect prior information \cite{jaynes1982rationale}. \\ 

Traditionally, Shannon entropy is used for the purpose of quantifying the uncertainty (i.e. entropy) of a derived probability distribution. Given a discrete random variable $X$ that can assume $N$ states and denoting by $p_i$ the probability for $X$ of being in state $i$, the formulation of entropy here is in the form of information, the summation of the probability \(p_i\) of observing $i$ out of $N$ possibilities multiplied by the logarithm of that probability, with base 2 being the traditional application: \(H = -\sum_{i=1}^{N}p_ilog(p_i)\) \cite{shannon1948mathematical}. Previous studies have shown that the maximization of (Shannon) entropy is useful because it allows an unbiased prediction. It has been also shown that Shannon entropy is the only functional form that satisfies `reasonable' properties, or axioms, from an inferential point of view\cite{shore1980axiomatic}. However, because many real-world systems do not adhere to such axiomatic structure (further outlined below), there has been considerable effort in relaxing those axioms, leading to a body of literature together called generalized measures of entropy (see also \cite{amigo2018brief}). While the relative merits of different entropy forms have been the subject of much debate in statistical inference, practical criteria for choosing among them are still lacking. This gap is particularly important now, as interest in applying MEP to complex systems continues to rise. Our contribution is to provide both a conceptual framework—by revisiting the Shore–Johnson axioms—and a practical workflow that indicates when generalized entropies are needed and how to implement them.\\ 

To provide the broader scope in the field of generalized entropies, in Section \ref{sec:axioms} we provide a conceptual overview of the axioms of Maximum Entropy as derived by Shore and Johnson \cite{shore1980axiomatic}. In Section \ref{sec:fam_entrop} we summarize various often-used quantitative measures of entropy. To show why a generalized treatment is necessary, Section \ref{sec:case_study} develops two concrete case studies—one drawn from economics, the other from ecology—that expose the practical hurdles of the standard approach. Readers who want a worked derivation can turn to Section \ref{sec:simple_model}, where we apply the generalized MEP to a minimal yet sufficiently rich model; this section may be skipped on a first reading. Finally, Section \ref{sec:discussion} discusses the theoretical and practical implications of the generalized framework and offers guidelines for selecting an appropriate entropy measure. Section \ref{sec:conclusion} concludes.\\

\begin{boxfloat}[!ht]
  \centering
  \begin{tcolorbox}[title=\textbf{Box 1  Classical Maximum-Entropy Recipe},
                   colback=blue!3!white,
                    colframe=blue!40,
                   width=\linewidth,
                   boxrule=0.4pt,
                   arc=2pt]
  \setlength{\abovedisplayskip}{6pt}
  \setlength{\belowdisplayskip}{6pt}

  \textbf{Problem.}\;
  Let $\Omega$ be a finite or countable set of micro-states $x$.
  For $m$ observables $f_i:\Omega\!\to\!\mathbb R$ we know their sample
  means
  \[
      \langle f_i\rangle \;=\; \sum_{x\in\Omega} p_x f_i(x) \;=\; F_i,
      \qquad i=1,\dots,m .
  \]
  Determine the \emph{least-biased}\footnote{%
    In the Shannon sense of preserving only the stated
    constraints (Shore–Johnson consistency).}
  probability distribution $P^{\*}=\{p_x\}$.

  \medskip\noindent
  \textbf{Solution.}\;
  Maximize Shannon entropy
  \[
      H(P)= -\sum_{x\in\Omega} p_x\ln p_x
  \]
  under the $(m\!+\!1)$ constraints
  $\sum_x p_x = 1$ and $\sum_x p_x f_i(x)=F_i$.
  Introduce Lagrange multipliers $\lambda_0,\lambda_1,\dots,\lambda_m$ and
  set the derivatives of the Lagrangian to zero:
  \[
      \mathcal L
      = -\sum_x p_x\ln p_x
        + \lambda_0\!\Bigl(\sum_x p_x-1\Bigr)
        + \sum_{i=1}^{m} \lambda_i\!\Bigl(\sum_x p_x f_i(x)-F_i\Bigr).
  \]
  Stationarity yields the \emph{exponential-family (Gibbs) form}
  \[
      p_x^{\*}
      = \frac{\exp\!\bigl(-\sum_{i=1}^m \lambda_i f_i(x)\bigr)}
             {Z(\boldsymbol{\lambda})},
      \quad
      Z(\boldsymbol{\lambda})
      = \sum_{x\in\Omega} \exp\!\bigl(-\sum_{i=1}^m
                                       \lambda_i f_i(x)\bigr).
  \]
  The multipliers are determined by the dual system
  \[
      F_i
      \;=\;
      -\,\frac{\partial \ln Z}{\partial \lambda_i},
      \qquad i=1,\dots,m,
  \]
  which can be solved numerically (e.g.\ Newton or gradient descent) or
  analytically for simple cases.

  \medskip\noindent
  \textbf{Takeaway.}\;
  Once the constraints $F_i$ and observables $f_i$ are specified,
  \emph{all} that remains is solving for
  $\boldsymbol{\lambda}$; the resulting $P^{\*}$ is automatically the
  maximum-entropy distribution.
  \end{tcolorbox}
  \captionsetup{labelformat=empty}
\caption{} 
\label{box:maxent} 
\end{boxfloat}
\clearpage
\section{The axioms from a practical point of view} \label{sec:axioms}

The when and why of generalized maximum entropy can be summarized in terms of assumptions that may or may not be violated, similar to how certain assumptions are violated for parametric statistical methods that then give the necessity of non-parametric approaches \cite{skilling1988axioms}. In the early developments, Jaynes already suggested that there exist such axiomatic conditions (although they were not named as such) that were later summarized and derived more thoroughly by Shore and Johnson as well as others (e.g. \cite{tsallis2015conceptual, jizba2019maximum, presse2013nonadditive}). The axioms to single out Shannon entropy for the Maximum Entropy Principle (MEP) can be described as the following:

\begin{itemize}
    \item Axiom 1: Uniqueness: the result should be unique.
    \item Axiom 2: Permutation invariance: state permutation should not matter.
    \item Axiom 3: Subset independence: it should not matter whether one treats disjoint subsets of system states in terms of separate conditional distributions or in terms of the full distribution.
    \item Axiom 4a: System independence (SI): It should not matter whether one accounts for independent constraints related to independent systems separately in terms of marginal distributions or in terms of full-system constraints and joint distributions.
    \item Axiom 5: Maximality: In absence of any prior information, a uniform distribution should be the solution.
\end{itemize}
In the original derivation by Shore and Johnson, however, the formal step used to derive the Shannon
functional corresponds to a \emph{stronger} condition that we will call
\textbf{strong system independence (SSI)}:
\begin{itemize}
    \item Axiom 4b: Strong System Independence (SSI): Whenever two
subsystems of a system are disjoint, we can treat the
subsystems in terms of independent distributions.
\end{itemize}
Whereas axioms 1, 2, 3 and 5 are generally considered by almost any form of entropy measure used, axiom 4a  potentially leads to distinct limitations in the use of MEP as statistical inference\cite{jizba2019maximum}. Shannon entropy, for example under independent constraints, leads to a joint distribution that factorizes. However, when using generalized entropy measures like Rényi entropy, even with independent constraints, the resulting joint distribution may not factorize, revealing the potential correlations between subsystems. This is further explained in more detail below, but the divergence between just these two examples is significant because it highlights the limitations of Shannon entropy in capturing potential dependencies that may arise from the intrinsic properties of the entropy measure itself.

Because these axioms may be unfamiliar, we (i) provide practical descriptions
of each axiom below, and (ii) include in Section \ref{sec:SSI} a concise Box that highlights the key
structural implication of \emph{SSI}: it enforces exponential growth of the phase space
with system size. In contrast, many complex systems exhibit sub- or super-exponential
growth, in which case Shannon entropy is not an adequate inference functional.
Generalized entropies become relevant in such cases, as they remain agnostic with
respect to the specific growth law of the phase space.

\medskip
\noindent\textit{Summary.} Shannon’s standard MEP corresponds to Axioms 1, 2, 3, 5 \emph{and}
the \emph{strong} version 4b (SSI). Some systems may violate axiom 4b. If one assumes only the \emph{weak} 4a (SI) rather than 4b,
a broader one-parameter family of generalized entropies is admissible; Shannon is recovered
as the special case where SSI (and exponential phase-space growth) happens to hold.
\subsection{Axiom 1: Uniqueness}
The first axiom states there should only be one single solution that 1) ensures the maximization of the information entropy of the derived distribution and 2) adheres to the given constraints from prior information. The intuitive explanation is straightforward: if the procedure of entropy maximization resulted in multiple solutions that could be considered equally likely, MEP would fail as a robust inference procedure as we would have no idea which solution is the best and provides the least biased result. When applying MEP to derive a probability distribution under known constraints from prior knowledge, this axiom ensures we will arrive at the same result – no matter what.\\

\subsection{Axiom 2: Permutation invariance}
The procedure of MEP can be considered symmetrical, this means that the labeling or ordering of the different states should not matter for the end result. In other words, the derived distribution from MEP should not change based on a particular order. This comes from the fact that no matter how things are labeled, the amount of information is the same. Of course, this is only valid for the states themselves, but that when changing the labeling of these states, the constraints connected to these labels should change in the same manner.\\

\subsection{Axiom 3: Subset independence}
This axiom deals with the fact that if we were to look at either the complete ensemble of states in terms of configuration space or at subsets of states comprising that complete ensemble should not matter. This means that either specifying the full system distribution or separate conditional distributions for states should not make a difference, the system itself is in some sense decomposable in its constituent elements.

\subsection{Axiom 4: System independence vs.\ Strong system independence} \label{sec:SSI}
When dealing with multiple systems that can be considered to be part of a larger system yet separated in for example time and/or space, this axiom addresses how to handle new information (i.e. constraints) that comes to light for one or more of the subsets but perhaps does not apply for all subsets. Note that this axiom deals with the independence of different \textit{systems} and not the independence of different subsets of states of one system (axiom 3), sometimes also referred to as strong system independence. If we are maximizing entropy for a specific distribution over a set of systems but we observe information that only applies for a subset of the entire ensemble of systems this should not influence the outcome that is not part of this subset. In other words, the derived probability distribution for those independent systems should remain unchanged. In quantitative terms this means that the information entropy of the two systems should be additive when both systems are considered in combination, hence the term strong additivity is sometimes used in place of strong independence. A slightly more mathematical explanation would be that either incorporating independent constraints for independent systems separately in terms of their marginal distributions, or combining them in a full system approach with constraints related to a joint-distribution should not matter for the end result.

What we have described above is the system independence axiom (SI, 4a). Strong system indepence (SSI, 4b) adds the stronger demand that the joint posterior distribution remains a
product measure under independent constraints. With Shannon entropy this implies strict
additivity $S_{\mathrm{Sh}}[AB]=S_{\mathrm{Sh}}[A]+S_{\mathrm{Sh}}[B]$ and, as shown in
Box~\ref{box_SSI}, enforces exponential phase–space growth. Generalized entropies
(e.g.\ Rényi/Tsallis) satisfy SI but \emph{need not} satisfy SSI; their MaxEnt joints can
exhibit correlations even when constraints are independent.
\\

\subsection{Axiom 5: Maximality}
The principle of maximum entropy is strongly related to the principle of indifference, dating back centuries at the beginning of probability thinking. This principles states that in the absence of alternative evidence, all possible outcomes should be considered equally likely. The maximization of entropy thus means that in the absence of any prior information a uniform distribution of probability over all possible states should be the solution. In slightly more mathematical terms, if we have system with $x_1$ to $x_n$ possible states but we have absolutely no prior information on how likely each would be, then the most uninformed prediction would simply be that $P(x_i) = 1/n$ for all $x$. In the spirit of the principle of indifference, the maximization of entropy ensures that we predict a distribution that only incorporates information we do have instead of introducing assumptions for which we have no evidence. \\

\begin{boxfloat}[!ht]
  \centering
  \begin{tcolorbox}[
      title=\textbf{Box 2  SSI enforces exponential growth of the phase space},
      colback=blue!3!white,
    colframe=blue!40,
      width=\linewidth,
      boxrule=0.45pt,
      arc=2pt]

\textbf{Setting (system size $N$ vs.\ sample length $k$).}
A single configuration of a size-$N$ system is $\sigma\in\Omega^{(N)}$ with probability $p_N(\sigma)$.
When we write a sequence $\Sigma^{(k)}=(\sigma^{(1)},\ldots,\sigma^{(k)})$, we mean $k$ \emph{independent} draws
from $p_N$ (i.i.d.\ configurations of length $N$).

\textbf{Independent constraints and SSI.}
Consider two disjoint subsystems $A$ and $B$ with state spaces $\Omega_A,\Omega_B$ and MaxEnt posteriors
$p_A^\star, p_B^\star$ obtained from constraints that \emph{do not couple} $A$ and $B$.
\emph{Strong System Independence (SSI)} requires the joint posterior to factorize:
\[
p_{AB}^\star(x,y)=p_A^\star(x)\,p_B^\star(y).
\]

\textbf{Shannon additivity for products.}
For any product measure,
\[
S_{\mathrm{Sh}}[p_{AB}^\star]
= S_{\mathrm{Sh}}[p_A^\star]+S_{\mathrm{Sh}}[p_B^\star],
\qquad
S_{\mathrm{Sh}}[p]:=-\sum p\log p.
\]

\textbf{Asymptotic Equipartition Property (AEP).}
Let $\Sigma^{(k)}$ be $k$ i.i.d.\ draws from $p_N$.
Then, by the law of large numbers applied to $-\log p_N(\sigma)$,
\[
-\tfrac{1}{k}\log p_N^{\otimes k}\!\big(\Sigma^{(k)}\big)
\;\xrightarrow[k\to\infty]{P}\; S_{\mathrm{Sh}}[p_N].
\]
Hence, for large $k$, almost all sequences have probability $\approx e^{-k S_{\mathrm{Sh}}[p_N]}$,
so the \emph{typical set} of sequences has size
\[
|A^{(k)}|\;\approx\; e^{k S_{\mathrm{Sh}}[p_N]}.
\]

This shows that $S_{\mathrm{Sh}}[p_N]$ measures the log of the effective number of configurations \emph{per draw}.
Accordingly, for a single configuration of size $N$ we define\footnote{If $p_N$ is uniform on $W(N)$ states, then $S_{\mathrm{Sh}}[p_N]=\log W(N)$ and

$W_{\mathrm{typ}}(N)=e^{S_{\mathrm{Sh}}[p_N]}=W(N)$.
The AEP shows this relationship extends to general (non-uniform) $p_N$:
$S$ still counts the logarithm of the \emph{effective} number of states.}
\[
W_{\mathrm{typ}}(N):=e^{S_{\mathrm{Sh}}[p_N]}.
\]

\textbf{From SSI to multiplicativity of typical counts.}
Applying AEP to $p_A^\star$, $p_B^\star$, and their product $p_{AB}^\star$, and using additivity:
\[
\begin{aligned}
W_{\mathrm{typ}}(N\!+\!M)
&= e^{S_{\mathrm{Sh}}[p_{AB}^\star]}
= e^{S_{\mathrm{Sh}}[p_A^\star]+S_{\mathrm{Sh}}[p_B^\star]} \\
&= W_{\mathrm{typ}}(N)\,W_{\mathrm{typ}}(M).
\end{aligned}
\]
Therefore $W_{\mathrm{typ}}$ satisfies the Cauchy multiplicative equation, whose positive solutions are exponential:
\[
W_{\mathrm{typ}}(N)=\mu^{\,N},\qquad \mu>1.
\]

\textbf{Takeaway.}
\[
\boxed{
\begin{aligned}
\text{SSI (factorization)}
&\;\Rightarrow\; S_{\mathrm{Sh}}\ \text{additivity} \\
&\;\Rightarrow\; \text{typical-set multiplicativity (AEP)} \\
&\;\Rightarrow\; W_{\mathrm{typ}}(N)=\mu^{\,N}.
\end{aligned}
}
\]

\textbf{Remark.}
The AEP always uses the Shannon entropy of the distribution $p_N$, regardless of which entropy functional (Shannon, R\'enyi, Tsallis, etc.) was used to derive $p_N^\star$. These are distinct roles of ``entropy.''

\end{tcolorbox}
  \captionsetup{labelformat=empty}
  \caption{} 
  \label{box_SSI}
\end{boxfloat}

\clearpage

\section{A family of Entropies}
\label{sec:fam_entrop}
Considering the above axioms, the only functional form of the entropy satisfying axioms 1, 2, 3, 5 and 4b is Shannon entropy \cite{shore1980axiomatic}. However, by relaxing Axiom 4b from Strong System Independence (formally used by Shore and Johnson in their proof) to System Independence (Axiom 4a), a one-parametric generalized family of entropies emerges as suitable for maximization in an inference procedure \cite{jizba2019maximum}.  

As mentioned in the previous section, the axiom of \textit{System Independence} states it should not matter whether one accounts for independent constraints related to independent systems separately in terms of marginal distributions or in terms of full-system constraints and joint distribution. Such relaxed version of the axiom states that the inference procedure has to remain coherent at all the scales MEP can be applied, but does not assume that independent information automatically translates into factorizable probabilities when both prior distributions and pieces of information are independent. Clearly, having no information about interaction encoded in constraints (i.e., having independent constraints) is not the same as having no correlations among systems. This axiom leads then to a generalized entropic family\cite{uffink1995can,jizba2019maximum}:
\begin{equation}
\label{eq:UJK}
H_q^{(f)}=f(\mathcal{U}_q(P))=f\bigg(\sum_i p_i^q\bigg)^{\frac{1}{1-q}}
\end{equation}
where $f$ is a monotonically increasing function. For clarity, when \textit{Strong System Independence} is enforced, the only possible $q$ is $q=1$, which results in the original Shannon entropy. Other notable entropies that are members of this family are R\'enyi entropy (corresponding to $f(x)=\ln x$) and Tsallis entropy (corresponding to $f(x)=\ln_qx= (x^{1-q}-1)/(1-q)$) (see below). Note that as $f$ is monotonically increasing, all these entropies share the same maximizing probability distribution. There exist many different functional forms of entropy and in the following we describe four often encountered forms for the practical user: Shannon entropy, relative Shannon Entropy, R\'enyi entropy and Tsallis entropy.

\subsection{Shannon entropy}
Shannon entropy, introduced in \cite{shannon1948mathematical}, is defined as:
\begin{equation}
    H(X) = -\sum_{i \in \chi} p_i \ln p_i.
\end{equation}
It quantifies the expected amount of information—or equivalently, the average level of uncertainty—contained in a probability distribution. It is additive for independent systems, i.e. $H(X,Y) = H(X) + H(Y)$ if $X$ and $Y$ are independent random variables. It is also worth to mention that the Shannon entropy of a uniform probability distribution $U$ is the Boltzmann entropy $H(U)=\ln \Omega$, where $\Omega$ is the volume of the state space (i.e. how many states the random variable can have).
\subsection{Relative Shannon entropy}
For two discrete probability distributions $p$ and $q$ over the same sample space $\chi$, the relative Shannon entropy (KL divergence) of $p$ relative to $q$ is defined as \cite{10.1214/aoms/1177729694}:
\begin{equation}
    D_{KL}(p||q)=\sum_{i\in \chi}p_i\ln \frac{p_i}{q_i}.
\end{equation}
Relative Shannon entropy measures how much a probability distribution diverges from a second one. In inference and statistical modeling, it is the proper function to maximize to infer the distribution $p$ under certain constraints, when the prior distribution is $q$. In this paper, we will focus in the standard maximum entropy approach, thus assuming the prior $q$ uniform.
\subsection{R\'enyi entropy}
R\'enyi entropy is defined as \cite{renyi1961measures}:
\begin{equation}
    R_q(X) \equiv H^{\ln}_q = \frac{1}{1-q}\ln\bigg(\sum_{i\in \chi}p_i^q\bigg).
\end{equation}
Note that different values of $q$ place different emphases on the tail behavior of the distribution. Because of this flexibility, R\'enyi entropy appears in diverse areas such as ecology, coding and various inference-based disciplines.\\

Interestingly, ecologists often evaluate and convert R\'enyi entropy to Hill numbers, sometimes called the 'effective number of species', via the exponential transform $^qD=\exp(R_q(X))$ which directly yields an intuitive idealized 'number of equally common species' interpretation. Varying $q$  produces a R\'enyi diversity profile that shows how a community’s diversity changes based on the relative emphasis on rare vs. abundant species \cite{hill1973diversity, jost2006entropy}. In coding theory, R\'enyi entropy emerges as compression limit when one wants to minimize the exponential average length of an encoded message, or, equivalently, the probability of buffer overflow \cite{campbell1965coding, somazzi2024nonlinear}. Note that from the point of view of entropy maximization, it has been proven that R\'enyi entropy is the only member of the family \eqref{eq:UJK} to retrieve consistency between the Maximum Entropy and the Maximum Likelihood principles \cite{somazzi2025learn}. Notable R\'enyi entropy properties are additivity for independent random variable (same as Shannon entropy), non-decreasing function of $q$, and that it recovers Shannon entropy: 
\begin{equation}
    \lim_{q\to1}R_q(X)=H(X).
\end{equation}
Finally, for a uniform distribution $U$, R\'enyi entropy is independent of $q$ and equal to Boltzmann entropy, i.e. $R_q(U) = \ln \Omega$.

\subsection{Tsallis entropy}
Tsallis entropy is defined as \cite{tsallis1988possible, tsallis2009introduction}:
\begin{equation}
    T_q(X) \equiv H^{\ln_q}_q = \frac{1}{q-1}\bigg(1-\sum_{i\in \chi}p_i^q\bigg).
\end{equation}
It was introduced to model systems where standard Boltzmann–Gibbs or Shannon approaches might fail, particularly those with strong correlations, long-range interactions, or fractal-like phase space structures (e.g., turbulence, astrophysical systems). Unlike Shannon entropy—additive for independent subsystems—Tsallis entropy typically exhibits non-extensive or '$q$-additive' behavior. When two systems are combined, their total entropy follows $T_q(X,Y) = T_q(X)+T_q(Y)+(1-q)T_q(X)T_q(Y)$. Also, while Tsallis entropy converges to Shannon entropy ($\lim_{q\to1}T_q(X)=H(X)$), when evaluated in the uniform distribution $U$ it does not retrieve Boltzmann entropy unless $q=1$. Instead, in general, $T_q(U)=(\Omega^{1-q}-1)/(1-q)$.


\section{Case studies}
\label{sec:case_study}
The entire discussion of the validity and applicability of the axioms described above goes beyond the scope this contribution. To illustrate the importance of their consideration, however, when we assume the axioms as explained above are in fact valid, we provide two case studies to exemplify the necessity of taking these into consideration. We specifically focus on axiom 4 in its strong and weak version (see section \ref{sec:axioms}) as we consider this to have the most practical relevance and provide an example for two different fields of study in which this approach has been previously used: economics and ecology.
\subsection{Economics}

For this example, we follow the formalism outlined in Caticha and Golan\cite{caticha2014entropic}. Consider the economies of two countries A and B that are separated by some trade barrier. Country A (resp. B) has agents labeled $a=1 \dots N_A$ (resp. $b=1 \dots N_B$). The goods are labeled $g=1 \dots G$. Let: i) $f_{ag}$ denote the amount of good $g$ produced by agent $a$; ii) $\chi_{agg'}$ be the amount of a good $g$ used as input for production of good $g'$ ; iii) $y_{ag}$ the amount used in general household consumption. The microstate of an economy is then a particular configuration of all inputs and goods:
\begin{equation}
\begin{aligned}
     \{\chi_{agg'}\}=x^{(A)}, \{y_{ag}\}=y^{(A)}\\
     \{\chi_{bgg'}\}=x^{(B)}, \{y_{bg}\}=y^{(B)}
\end{aligned}
\end{equation}
In Cathica and Golan\cite{caticha2014entropic}, the probability distribution over those microstates for a country A (and similarly for country B) is obtained by maximizing the Shannon entropy while satisfying the following constraints on production and consumption:
\begin{enumerate}
\item \(\langle \sum_{a g'} \chi_{a g g'} \rangle = X^{(A)}_g\) \quad (average usage of good \(g\) in production for \(A\)),
\item \(\langle \sum_{a} y_{a g} \rangle = Y^{(A)}_g\) \quad (average consumption of good \(g\) for \(A\)),
\item \(\langle \sum_{a} f_{a g} \rangle = F^{(A)}_g\) \quad (average production of good \(g\) for \(A\)),
\end{enumerate}
Because we assume a trade barrier, all consumed goods must have been produced in the focal country. Therefore, an additional constraint says:
\begin{equation}
    F^{(A)}_g=X^{(A)}_g+Y^{(A)}_g
\end{equation}
reflecting that a country must ``use up'' the goods it produces either in further production or in consumption. For country \(B\), a similar relationship holds.

Now, if we treat these two countries as separate systems and impose only independent constraints for each one, the standard Shannon-entropy maximization will lead to factorized distributions:
\[
    p\bigl(x^{(A)}, y^{(A)}, x^{(B)}, y^{(B)}\bigr) 
    \;=\;
    p\bigl(x^{(A)}, y^{(A)}\bigr)\;p\bigl(x^{(B)}, y^{(B)}\bigr).
\]
This factorization is a direct consequence of the strong system independence (SSI) axiom in Shannon's framework (see section \ref{sec:SSI}).

However, real-world economies frequently display correlations that do not vanish merely because trade barriers or other constraints appear to isolate them. For example, both countries still share the same planet's finite resources. A disruption in one region's resource extraction may ripple into other regions. Geopolitical events (wars, pandemics, or financial market crashes) can simultaneously affect production and consumption in each country, creating correlations even when there is no active trade.
Finally, indirect spillover effects (through innovation, global supply chains, environmental impacts) can correlate the two economies, even if those interactions are not captured by explicit constraints.

In other words, the absence of explicit ``cross-country'' constraints in the optimization does not necessarily mean the two subsystems are truly independent; it simply means one has not explicitly modeled their interdependence. By using Shannon entropy subject only to separate constraints, one unavoidably loses any hidden correlation---Shannon's strong system independence enforces factorization.

On the other hand, generalized entropies (e.g., R\'enyi, Tsallis, or other non-Shannon forms) relax the assumption of strong system independence and thus allow for the possibility that correlations might exist or emerge naturally, even if the constraints do not explicitly couple the countries.

Hence, if one has a strong reason to suspect subtle interdependencies---such as resource limits, global shocks, or other correlated phenomena---but can only enforce ``independent'' constraints in your data-fitting procedure, a generalized maximum entropy framework might be more faithful to the underlying reality. It can accommodate correlations that remain hidden under a purely Shannon-based treatment.

In practical economic modeling, one often lacks complete or precise information on all the ways countries might influence each other. For instance, one might not have exact data on resource constraints, spillover effects, or global events. Nonetheless, if strong system independence is an unrealistic assumption,  generalized maximum entropy is a more careful approach. Even when only separate constraints can be reliably enforced, these generalized forms retain enough flexibility to capture correlations that Shannon's factorization would discard.

This example demonstrates the broader conceptual point brought up by Jizba and Korbel\cite{jizba2019maximum}, i.e. lacking explicit information about interactions (i.e., using independent constraints) does not necessarily imply an absence of correlations between subsystems.

\subsection{Ecology}
In ecology, a question often asked is how many species there are in a given habitat and how this can be estimated from limited sampling. This case study considers this question and shows the consequence of violating Strong System Independence (SSI). First, consider a forest where two separate inventories have been carried out in location A and B in order to estimate the diversity of the entire forest. Location A (resp. B) has species labeled $s_a=1 \dots S_A$ (resp. $s_b=1 \dots S_B$). The number of individuals for each species in either location are labeled $n=1 \dots N$ with the relative abundances given by $p_i=S_{(A)i}/n_i$. We assume there is some overlap between the species found in either location but that the abundances vary and that each site also has unique species.\\

If we assume that the two plots are independent, following the SSI axiom 4b, the total entropy of the forest based on these samples would simply be the joint entropy given by the sum of their individual entropies (\(H_{\text{joint}} = H_A + H_B)\).  System independence fails, however, if the species abundance distributions in location A and B are not independent, for example by shared species or shared environmental characteristics of which we might not even be aware. In other words, knowing something about location A would already give information on what to expect for location B. In this case, we should account for the overlap and so-called shared information between the two, defined as: \(H_{joint}\ = H_A + H_B - I(A; B)\). The last term in this equation is termed mutual information, representing the degree of dependence between the two systems. The mutual information between two systems can be stated as the marginal entropy of the A minus the conditional entropy of A given what we know about B:

\begin{equation}
\begin{split}
H(A; B) & = H(A) - H(A|B) \\
 & = H(B) - H(B|A) \\
 & = H(A) + H(B) - I(A,B)
\end{split}
\end{equation}

I(A,B) here stands for the shared or mutual information between plots A and B, given by $\sum_{(i,j)} P(i,j) log \frac{P(i,j)}{P(i)P(j)}$, $H(A)$ and $H(B)$ respectively for the separate marginal entropies of plot A and B and $H(A|B)$ or $H(B|A)$ for the conditional entropy of either A or B taking into account B or A and finally H(A,B) for the joint entropy of the two plots taken together.\\

From this set of equations, it is clear that if there is no shared information (i.e. there is no conditional entropy from knowing something about A or B that concerns either of the two) the entropy of the forest as a whole is indeed the two separate entropies combined. If there is, however, shared information, then the conditional entropy is positive and the entropies no longer factorize. The total entropy of the system should then be corrected for the mutual information.\\

A simple example to illustrate the consequence of violating the Strong System Independence axiom is when we try to estimate the diversity of the forest by calculating the so-called \textit{effective} number of species. Introduced by MacArthur in 1965, for Shannon entropy measures of diversity, this is given by $S_{eff} = e^H$. If we assume hypothetically that for location A, $H(A) = - \sum_{i} p_A(i) \log p_A(i) = 1.8$ and for location B, $H(B) = - \sum_{i} p_B(i) \log p_B(i) = 1.4$, two scenarios are possible: either there is no mutual information, $I(A,B) = 0$ or that $I(A,B) > 0$. In the former case, the effective number of species is then given by $S_{eff} = e^{H(A)+H(B)}$, while in the latter $S_{eff} = e^{H(A)+H(B)-I(A,B)}$. If we assume that there are several shared species between the two locations, for example because these are very dominant in the regional area the mutual information could be positive, e.g. $I(A,B)$ could be .34. If we then take these numbers, it is clear that we would overestimate the number of effective species if we did not take into account these dependencies:

\begin{equation}
\begin{split}
uncorrected: S_{eff} = e^{1.4+1.8} \approx 25 \\
corrected: S_{eff} = e^{(1.4+1.8)-.34} \approx 18 \\
\end{split}
\end{equation}

Thus, if the overlap is ignored and system independence is assumed, the combined entropy (and therefore diversity) of the forest will be overestimated as one does not account for the shared information between the two system. Once again, even if we lack information about the interactions between these two plots, it does not imply absence of correlation. By applying the maximization of entropy approach in the \emph{Shannon} functional form in presence of independent constraints on these two plots, the joint probability of the two automatically factorizes, leading to $I(A,B)=0$ and a possible overestimation of effective species in this case.

A generalized entropy approach however can handle violations of strong system independence by using other functional forms of the entropy maximization, like the R\'enyi or Tsallis entropy which offer a more flexible framework to account for potential interdependencies, e.g. less or more emphasis on dominant species for examples that are shared between the two plots.

An important note of consideration is that we should be careful to distinguish between the entropy form used to infer the probability distribution of, for example, species abundance or estimations of effective species diversity and the entropy form of Shannon to as a diversity index. In the face of possible correlations, one should use generalized entropy to infer the probability distribution for the first and then, if interested in Shannon diversity, one can safely use the Shannon entropy to quantify it for both samples simultaneously assuming they are reflecting of the whole system.

\section{Simple model}
\label{sec:simple_model}
This section introduces a minimal model that violates strong system independence. A signature of this violation is the sub-exponential scaling of the phase space (number of feasible configurations) as a function of the number of degrees of freedom. Intuitively, this indicates the presence of strong correlations among the constituents of the system that make the strong system independence axiom non applicable (see Section \ref{sec:discussion}).

Consider a one-dimensional chain consisting of $N$ binary correlated random variables (bits), taking values $s_i\in \{0,1\}$. Although abstract, binary random variables are a general description that can capture many applications. For example, the ``chain" can be considered as a list of species found in a sample where the presence or absence of a species (i.e. 1 or 0) is or is not dependent on the other species presence or absence (i.e. the other variables in the chain), conditional on whether we are dealing with a system satisfying strong system independence or not. Hereafter, we refer to the chain of binary random variables as a general form.

Taking inspiration from Ruseckas~\cite{ruseckas2015probabilistic}, we consider that bits next to each other have almost always the same value, but there are $d$ cases (``flips") where the next bit has a different value. For example, if $N=10$ and $d=2$, the sequence ``0011111100" is allowed, with flips in positions 2 and 8. This rule effectively decreases the number of allowed configurations; in particular, while an unconstrained chain of bits' phase space scales exponentially with respect to $N$, in this case it scales as a power of $N$, precisely as $\Omega_d(N) \sim (2/d!) N^d$.

By calling $M=\sum_i^N s_i$ the sum of the chain (i.e. the total number of species corresponding to the value $s_i=1$), from \cite{ruseckas2015probabilistic} we also know the number of configurations leading to a given sum $M$, for $N>>d$:
\begin{equation}
    \Omega_d(N,M) = \frac{N^{d-1}}{\kappa(d)}\Big(1-4\frac{(M-N/2)^2}{N^2}\Big)^{\lfloor \frac{d-1}{2}\rfloor}
\end{equation}
where $\kappa(d) = d^{d-2}\lfloor \frac{d}{2}\rfloor!\lfloor \frac{d-1}{2}\rfloor!$ and $\lfloor \rfloor$ is the floor function.

\begin{figure}[ht]
    \centering
    \begin{subfigure}[t]{0.48\linewidth}
        \centering
        \includegraphics[width=\linewidth]{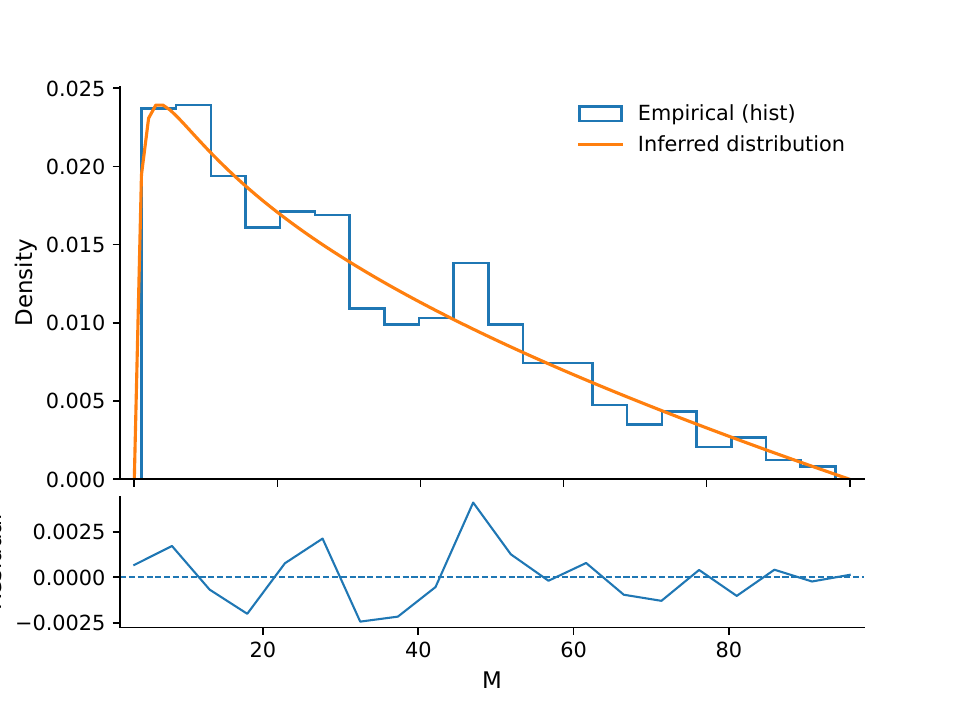}
        \caption{}
        \label{fig:hist_infer}
    \end{subfigure}
    \hfill
    \begin{subfigure}[t]{0.48\linewidth}
        \centering
        \includegraphics[width=\linewidth]{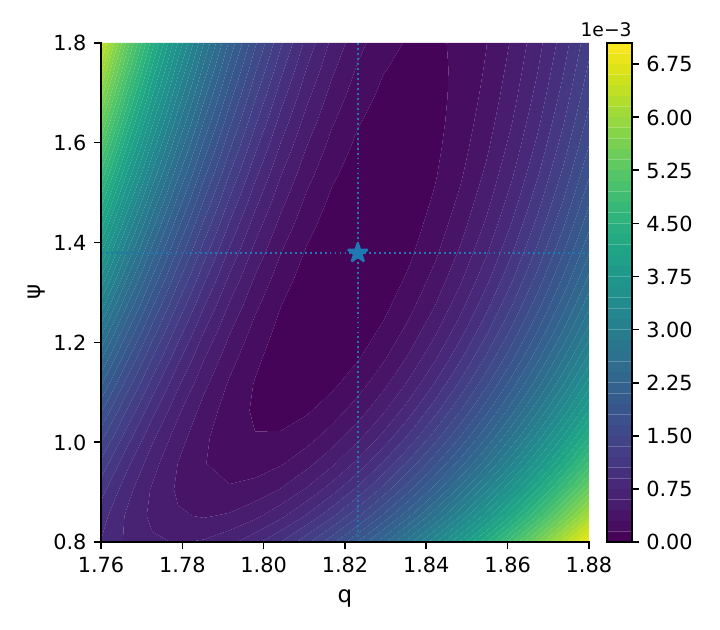}
        \caption{}
        \label{fig:lik_profile}
    \end{subfigure}
    \caption{(a) Histogram of observed $\Vec{M}^*$, together with inferred probability distribution $p(M|q^*,\psi^*)$. (b) contour plot of the 
log-likelihood surface, expressed as the difference 
$\ell(\hat{q},\hat{\psi})-\ell(q,\psi)$ between the maximum 
value (at the MLE $(\hat{q},\hat{\psi})$) and the value at 
each $(q,\psi)$.}
    \label{fig:hist_p_lik}
\end{figure}

By considering a Generalized Maximum Entropy approach imposing the $q$-average of $M$ as constraint, as described in \cite{somazzi2025learn}, one gets that the probability of a given configuration $c=\{s_1,\dots,s_N\}$ is:
\begin{equation}
    p(c) = \frac{1}{Z_q(\psi)}\Big(1-(1-q)\psi M(c) \Big)_+^\frac{1}{1-q}
\end{equation}
where we have used the notation $(x)_+^a \equiv 0$ if $x<0$, while
$(x)_+^a \equiv x^a$ otherwise. $Z_q(\psi)$ is the normalization factor, also called partition function. It can be re-written as:
\begin{equation}
    Z_q(\psi) = \sum_c \Big(1-(1-q)\psi M(c) \Big)_+^\frac{1}{1-q} = \sum_{M=0}^N \Omega_d(N,M) \Big(1-(1-q)\psi M \Big)_+^\frac{1}{1-q}.
\end{equation}
The power-law scaling of $\Omega_d(N,M)$ allows for non-trivial distributions, and for the analytical solution of $Z_q(\psi)$ (and therefore of other quantities such as averages like $\langle M \rangle$) for some parameters' values. 
However, from the point of view of statistical inference, we will focus here on the estimation of parameters from observations, as detailed in \cite{somazzi2025learn} (see Box \ref{box_learn} for a quick recap). Given a series of $K$ independent observations\footnote{We imagine that such observations come from controlled repeated experiments, meaning that we know they are independent of each other, and thus their joint probability factorizes into the product of the marginals.} $\Vec{M^*}=\{M^*_1, \dots, M^*_K\}$, we would like to maximize the log-likelihood of our data with respect to the parameters $q$ and $\psi$:
\begin{equation}
    \nabla_{(q^*,\psi^*)}\log \prod _{k=1}^K p(c_{M^*_k}|q,\psi)=\Vec{0}.
\end{equation}
The selected parameters $(q^*,\psi^*)$ are the best unbiased estimators according to the principles of maximum likelihood inference. 
We run a simple numerical experiment, considering a chain of length $N=100$ and flips $d=4$. We sampled $\Vec{M}^*=\{M^*_1,\dots,M^*_{1000}\}$ values to learn our parameters, according to a power law distribution.
The resulting values are $(q^*,\psi^*) \approx (1.821, 1.337)$, meaning that our inference procedure suggests a different entropy and thus a non-exponential distribution ($q\neq 1$).
Figure \ref{fig:hist_infer} shows the histogram of the observed $\Vec{M}^*$ and the overlapped inferred distribution $p(M|q^*,\psi^*)$. Figure \ref{fig:lik_profile} shows a contour plot of the 
log-likelihood surface, expressed as the difference 
$\ell(\hat{q},\hat{\psi})-\ell(q,\psi)$ between the maximum 
value (at the MLE $(\hat{q},\hat{\psi})$) and the value at 
each $(q,\psi)$. This re-centering sets the optimum at zero 
and highlights how quickly the likelihood declines away 
from the maximum, a standard approach that also connects 
directly to likelihood-ratio confidence regions.

The example above shows that by using a generalized maximum entropy approach, together with a maximum likelihood estimation procedure\cite{somazzi2025learn}, the ``correct'' entropy emerges from the data. In this case, the power law scaling of the phase space indicated that strong system independence is violated and Shannon entropy would be inapplicable. If it were not the case, the maximum-likelihood procedure --- provided sufficient amount of data --- would retrieve the Shannon case ($q=1$).

\begin{boxfloat}[!ht]
  \centering
  \begin{tcolorbox}[
      title=\textbf{Box 3  Learning the Entropic Parameter from Data},
      colback=blue!3!white,
    colframe=blue!40,
      width=\linewidth,
      boxrule=0.45pt,
      arc=2pt]

  \textbf{Problem.}\;
  Let the system take micro-states \(x\in\Omega\).
  Each state carries an observable (or vector of observables) \(C(x)\).
  We record \(M\) i.i.d. realizations
  \(\mathcal D=\{x^{(m)}\}_{m=1}^{M}\) with values
  \(C^{(m)}:=C\bigl(x^{(m)}\bigr)\).
  
  Determine the \emph{least-biased} probability distribution $P^{\*}=\{p_x\}$.
  
  \medskip\noindent
  \textbf{Solution.}\;
  Maximizing a Rényi entropy with the
  \(q\)-mean constraint yields 
  \[
     p_q(x,\psi)=
     \frac{\bigl[1-(1-q)\,\psi\!\cdot\!C(x)\bigr]_{+}^{1/(1-q)}}%
          {Z_q(\psi)} ,
  \]
  \[
     Z_q(\psi)=\sum_{x\in\Omega}
               \bigl[1-(1-q)\,\psi\!\cdot\!C(x)\bigr]_{+}^{1/(1-q)} .
  \]

  \medskip\noindent
  We need to find the pair \((q^\ast,\psi^\ast)\) that maximizes the sample log-likelihood.

  \medskip
  \rule{\linewidth}{0.4pt}

  \noindent\textbf{Step 1 – optimise \(\psi\) for fixed \(q\).}
  For each trial \(q\) maximise
  \[
     \bar\ell_q(\psi)=
     \frac1M\sum_{m=1}^{M}\!
        \ln p_q\bigl(x^{(m)},\psi\bigr)
  \]
  to obtain \(\psi_q^\ast\).
  Stationarity gives the \(q\)-mean matching rule
  \[
     \sum_{x\in\Omega}\!C(x)\,p_q^{\,q}(x,\psi_q^\ast)
     =\frac1M\sum_{m=1}^{M}
        C^{(m)}\,p_q^{\,q-1}
        \!\bigl(x^{(m)},\psi_q^\ast\bigr).
  \]

  \medskip\noindent
  \textbf{Step 2 – select \(q^\ast\).}
  Evaluate the partially maximised score
  \(\bar\ell_q^{\max}:=\bar\ell_q(\psi_q^\ast)\)
  on a grid (or via one-dimensional optimisation) and choose
  \[
       q^\ast=\arg\max_{q}\bar\ell_q^{\max},
       \qquad
       \psi^\ast:=\psi_{q^\ast}^\ast.
  \]
  \noindent\textbf{Practical suggestions.}
  \begin{itemize}
    \item Use Newton or L-BFGS for \(\psi_q^\ast\)\,—one solve per \(q\);  
          then Brent or log-parabola search on \(\bar\ell_q^{\max}\).
    \item \(q^\ast=1\) collapses to the classical recipe in Box 1;  
          \(q^\ast\neq1\) flags violation of strong system independence.
    \item Heavy-tailed samples need larger \(M\) before \(q^\ast\) stabilises.
  \end{itemize}

  \end{tcolorbox}
  \captionsetup{labelformat=empty}
  \caption{} 
  \label{box_learn}
\end{boxfloat}
\clearpage
\section{Discussion and future outlook}
\label{sec:discussion}
The traditional use of the functional form introduced by Shannon for the maximization procedure of entropy has had a long history and many applications in a diverse array of disciplines. Simultaneously there has been a considerable debate on the applicability of specifically the functional form of the Shannon entropy given the underlying assumptions or axioms associated with it. Here we have discussed these axioms, their implications and the challenges they bring. Most of the axioms can be considered universal, meaning they are valid for any form of entropy (e.g. Shannon, Rényi, Tsallis etc.). Some, however, offer a more complicated discussion. Given the importance of these axioms for the calculations of entropy, there is a growing body of research that explores the use of more generalized entropy functions —such as Rényi or Tsallis entropy— as flexible alternatives for situations that violate some of the foundational assumptions of classical entropy. Much of it is, however, primarily of a theoretical nature and a more intuitive explanation was missing. Here we offered a conceptual overview of the axioms with a specific emphasis on the axiom of system independence to illustrate some of the consequences and hopefully bring the concept of generalized entropy to a wider audience.
The major challenge we consider is that the classic Shore–Johnson axiomatic derivation of the Maximum Entropy Principle (MEP) assumes implicitly the strong system independence (SSI) axiom: whenever two subsystems are disjoint, one may impose constraints on each separately or on the joint system interchangeably. Imposing SSI forces the entropic functional to factorize additively and thereby uniquely singles out Shannon entropy as the only viable inference measure. There are, however, many alternatives to the functional form of the entropy maximization and when we take a more general look we see that Shannon entropy is simply part of a larger (one-parameter) family of entropies. In fact, if SSI is not imposed, the Shore and Johnson consistency still holds for the broader Uffink–Jizba–Korbel class of functionals for the maximization of entropy given in its general form by,

    \[
      U_q(P)\;\propto\;\Bigl(\sum_i p_i^q\Bigr)^{1/(1-q)},
    \]
modulo any monotonic transformation. In other words, relaxing SSI yields a range of one‐parameter entropies, with Shannon as the $q\to1$ limit. The question that arises is then for any practical applications: when do we expect the SSI criterion to fail?\\

\textit{Phase‐space growth of a complex system}.  A concrete criterion for SSI’s validity is the so-called scaling of the phase‐space volume $W(N)$ with system size $N$. The phase-space volume of a system here is defined as the potential number of microstates of the system, i.e. all the possible configurations of the constituent elements of a system given certain macroscopical constraints. With increasing size of a system, the potential configuration of the system grows, but what is important is the way it grows. For independent systems, the phase space of the system $W$ given $N$ grows exponentially. A simple example shows this to be true: if a subsystem can be in either two states, A or B, for each added entity, the possible number of configurations doubles. For $N=1$ there are two possibilities, either the entire system is in state A or B, for $N = 2$ this becomes four possibilities (AA, AB, BB, BA), for $N= 3$ this becomes 8 etc. In other words, from this the relationship with additivity as outlined above becomes clear as the exponential growth of the phase space system means that $\log(W(N))$ scales linearly with $N$. In the case of failure of independence, the issue is not the presence of correlations per se. The core problem is that $W(N)$ is no longer an extensive property (i.e. a property of which the magnitude is dependent on the size of the sample and additive for subsystems). The potential number of configurations of the system might be limited because of this correlation. In other words, given the same example, when for example for $N=1$ we know the system to be in state A this might impose constraints that for $N=2$, the system must either be AB, BA or AA (i.e. it must include an A). This means the growth of the phase-space occurs in a non-exponential manner and could be more like a power-law. The assumptions of strong system independence, and thus additivity in log space no longer apply. In more mathematical terms, if

\[
      W(N)\sim e^{c_0 N},
    \]
then subsystems remain effectively independent and Shannon ($q=1$) applies. Whenever $W(N)$ deviates, sub‐exponentially (power‐law) or super‐exponentially, SSI fails and one should instead employ a generalized entropy with $q\neq1$. In other words, the way $W(N)$ scales with increasing $N$ gives us information on how the system behaves and if the SJ-axiom of SSI holds. How can we incorporate this in our analyses?\\

\textit{A scientists guide to entropies}.  Given these considerations, we provide a step-by-step procedure that we believe should be considered by any user of the maximization of entropy to ensure the proper use of this otherwise powerful framework:

\begin{enumerate}
\item Explore the dimensionality of the system under study, i.e. what constitutes the potential phase space of the system, what are the underlying structures of the probability distributions over which maximization of entropy will be applied? In this step it is important to get a more conceptual perspective of the potential caveats.
\item Analytically diagnose the scaling of $W(N)$ (e.g. via sample‐space enumeration, fitting to various functional forms of scaling or empirical partition‐function fits).
\begin{enumerate}
\item If $W(N)\propto e^{O(N)}$, stick with Shannon entropy.
\item Else, adopt the generalized entropy $U_q(P)$ and determine the data‐driven value of $q$ by for example maximum‐likelihood estimation: treating $q$ as an extra parameter in the MEP “learns” $q$ from the data, reducing to Shannon exactly when the system truly exhibits exponential scaling (see Box \ref{box_learn} and \cite{somazzi2025learn}).
\end{enumerate}
\item Visualize different outcomes of using various forms of the entropy families by adjusting the q-parameter and general entropy profiles.
\item Decide on entropy measure and outline considerations in methodology for reproducibility and accountability. 
\end{enumerate}

\section{Conclusion}
\label{sec:conclusion}
Although our analyses of (complex) systems can be greatly aided by tools from information theory it also brings with it the requirement of more abstract notions of how we think of these systems. In essence, it requires a more probabilistic point of view and with it the associated rules of engagement related to calculations of probabilities. From an applied standpoint, nothing fundamentally changes about the choice of constraints when moving from Shannon to a generalized entropy: one still conditions on expectation values, solves via Lagrange multipliers, and only afterwards determines those multipliers (and q itself) by fitting to observed averages. There is, however, a deeper consideration necessary that we have emphasized here that we believe is critical if we are to derive meaningful conclusions. We hope this conceptual overview provides some of the insights necessary for practitioners to consider the limitations of these methods but also their power to analyze real-world systems.
\section*{Acknowledgements}
G.M.F. conducted most of this research while affiliated with the Department of Ecology and Evolutionary Biology at Princeton University. G.M.F. acknowledges support from the Swiss National Science Foundation (grant P500PS-211064), a gift from William H. Miller III, and the Schmidt DataX Fund at Princeton University, made possible through a major gift from Schmidt Futures.\\
A.S. acknowledges support by the Dutch Econophysics Foundation (Stichting Econophysics Leiden, the Netherlands) and by the European Union - NextGenerationEU - National Recovery and Resilience Plan (Piano Nazionale di Ripresa e Resilienza, PNRR), projects ``SoBigData.it - Strengthening the Italian RI for Social Mining and Big Data Analytics'', Grant IR0000013 (n. 3264, 28/12/2021) (\url{https://pnrr.sobigdata.it/}), and ``Reconstruction, Resilience and Recovery of Socio-Economic Networks'' RECON-NET EP\_FAIR\_005 - PE0000013 ``FAIR'' - PNRR M4C2 Investment 1.3. He also acknowledges support by the European Union under the scheme HORIZON-INFRA-2021-DEV-02-01 – Preparatory phase of new ESFRI research infrastructure projects, Grant Agreement n.101079043, “SoBigData RI PPP: SoBigData RI Preparatory Phase Project”.

\section*{Data availability}
This study did not generate new empirical datasets. 
The code used for the illustrative example is available at 
\url{https://github.com/Peppe95/Generalized_Maximum_entropy_when_why}.


\bibliographystyle{unsrt}
\bibliography{references}
\end{document}